# Open-ocean-interior moored sensor turbulence estimates, below a Meddy

**by Hans van Haren**

Royal Netherlands Institute for Sea Research (NIOZ) and Utrecht University, P.O. Box 59, 1790  AB Den Burg, the Netherlands.
*e-mail: hans.van.haren@nioz.nl

ABSTRACT

A one-year time series of moored high-resolution temperature T-sensor data from 1455±65 m depth on a 3900 m long line in about 5300 m of water in the NE-Atlantic Canary Basin are dominated by salinity (over-)compensated intrusions arising from the effects of Mediterranean outflow waters, which are commonly organized as 'Meddies'. During the passage of a Meddy-core above the T-sensors, no intrusions were observed, thereby making it possible to use the temperature records to quantify turbulence parameters. Mean turbulence values confirm previous shipborne profiler estimates of $O(10^{-10})$ m$^2$ s$^{-3}$ and $O(10^{-5})$ m$^2$ s$^{-1}$ for dissipation rate and eddy diffusivity, respectively. The present data show that these ocean-interior turbulence estimates are from short-lived (<0.5 h) rather intense overturning cells with vertical scales of <5 m. Because the turbulence inertial subrange is found to extend into the internal wave band, the overturns are predominantly driven by shear associated with inertial currents. Kinetic energy, current shear and temperature variance peak at sub-inertial frequencies during the Meddy passage, suggesting wave trapping in the warm anti-cyclonic eddy and/or weakly stratified layers. The observations further show that internal wave displacements are coherent over vertical scales of up to 40 m during the presence of the Meddy compared with vertical coherence scales of <25 m during the more common no-Meddy conditions of double diffusion intrusions.

## 1. Introduction

The open-ocean interior away from surface and (sloping) bottom boundaries is considered a well-stratified and generally weakly turbulent domain, even though it supports substantial motions, e.g., by internal waves. Shipborne profiling instruments have confirmed this picture of the ocean interior and turbulence parameter values are 100 to 1000 times smaller than near boundaries (e.g., Gregg, 1989; Polzin et al., 1997). See Table 1 for some mean values, noting that short-term turbulence values vary over four orders of magnitude in the ocean. Yet, the ocean-interior turbulent exchange or 'diapycnal mixing' is still 10 to 100 times larger than that of molecular diffusion. Several processes have been suggested to generate turbulence in the ocean interior, ranging from episodic internal wave breaking to double diffusion. Thus far long-term moored high-resolution observations have seldom shown such turbulence processes.

A mooring array of densely spaced high-resolution temperature (T) sensors could potentially investigate the evolution of ocean-interior turbulence, under particular conditions. Based on the overturning method by Thorpe (1977), salinity (over-)compensated intrusions however negate quantification of turbulent parameters from such T-data. This is because a non-tight temperature-density relationship leads to our inability of using T-data as a proxy for density variations. Intrusions show irregular isotherms that falsely suggest vigorous mixing. It has been proven by Oakey (1988) that large temperature microstructure variability does not necessarily imply large turbulence dissipation rates. This conclusion was drawn after analysis of extensive shipborne turbulence microstructure-shear and -temperature observations from a 'Meddy', an underwater eddy-lens of Mediterranean water in the North-East Atlantic Ocean.

The Mediterranean waters reach far into the NE-Atlantic (Fig. 1a). Because of their relatively large density with respect to Atlantic surface waters they are found about 1000 m deeper than the 280 m water depth of the Camarinal Sill in the Strait of Gibraltar (Antonov et al., 2006). In detail however, the salinity excess is not organized as depicted in the annual mean distribution of (Fig. 1a), but it is concentrated in much smaller sub-surface mesoscale Meddies that have diameters of 10-100 km. These relatively warm, salty lenses live about two years and are pulsed into the NE-Atlantic in a corridor between the Azores and Canary fronts (Sangrà et al., 2009).

Meddies have been studied extensively in situ in the 1980s and 1990s (e.g., Armi et al., 1988; Hebert et al., 1990; Ruddick, 1992; Pingree and Le Cann, 1993). Ocean dynamics models demonstrated surface effects of Meddies which resulted in an eddy-dominated NE-Atlantic (e.g., Spall, 1990; Bashmachnikov et al., 2009; Sangrà et al., 2009). The in situ observational studies demonstrated a 'pristine' low-turbulent core of the Meddy with salinity increasing with depth (Ruddick et al., 2010). Above and below the core double diffusion processes were dominant. Below the core, double diffusion was dominated by salt fingering such that both salinity and temperature decrease with depth. The steepest gradients in temperature and salinity were found directly under the core, with less steep gradients under the Meddy-sides (Hebert et al., 1990). On the lateral Meddy-sides intrusions were found. Intrusions are strongly layered phenomena of alternating large and small gradients in temperature and salinity of which the precise drivers are not known (Meunier et al., 2015). As will be demonstrated here from yearlong moored high-resolution temperature T-data, intrusions are ubiquitous in the interior of the NE-Atlantic Canary Basin.

Reanalysis of these moored T-data shows that underneath a Meddy-core intrusions are indeed absent. There, a different and tighter temperature-density relationship is found than in surrounding waters and T-data may be analyzed for their turbulent overturning. It is hypothesized that such lower Meddy-cores represent the ocean-interior turbulence far from boundaries, in an environment of extensive near-linear internal wave propagation and potential episodic breaking. After all, Meddies may interact with but do not hamper the passage of internal waves that are ubiquitous throughout the stratified ocean.

The mooring was part of an extensive program on near-inertial internal wave studies in NE-Atlantic Basins. As an additional aim, the present 1.5 year-long mooring was meant to monitor ocean-interior turbulence from T-sensor data where episodic internal wave breaking was expected. The top-buoy was located around 1400 m depth to avoid fishing hazards, which had the expected advantage of being deeper than the Mediterranean waters so that salinity (over-)compensating intrusions were avoided. The additional aim was abandoned after finding the dominance of inversions in T-sensor data due to intrusions. However, it led to an investigation of particular ocean-interior internal wave spectral details (van Haren and Gostiaux, 2009).

In the present paper, the focus is on three weeks of a 100-day period when a Meddy passes over the T-sensors. During this passage no T-inversions reminiscent of intrusions occur, which renders a completely different temperature variance compared with the data before and after this 100-day period. While the structure of observations clearly suggests a tight temperature-density relationship just below the quiescent Meddy-core, direct evidence from shipborne Conductivity-Temperature-Depth CTD-data near the mooring is not available during the Meddy passage. CTD-profiles were only obtained during the mooring deployment/recovery cruises, once every 1.5 years. The main questions are how the temperature distribution differs in and around a Meddy, how turbulence levels vary in a Meddy and what process generates Meddy-core turbulence. A better understanding is anticipated of the variability of turbulent diffusion in the ocean-interior and its association with the permanent presence of internal wave motions.

## 2. Materials and methods

Between 10 June 2006 (day 160) and 22 November 2007 (day 324+365), a 3900 m long mooring was deployed in the open Canary Basin, NE-Atlantic. The coordinates were 33° 00.0′N, 22° 04.8′E, H = 5274 m water depth. The top-buoy held a low-resolution downward looking 75 kHz Teledyne-RDI LongRanger acoustic Doppler current profiler ADCP. It sampled 60 vertical bins of 10 m, with a horizontal averaging scale between 20 and 420 m depending on the vertical distance and due to the 20° slant angle of the acoustic beams to the vertical. From 15 m below it, 54 high-resolution NIOZ3 T-sensors were taped to the mooring cable at 2.5 m intervals. The highest sensor was at 1389 m and the deepest at 1522.5 m. Here, we focus on the first year of data when only 4 T-sensors showed electronic noise, calibration, or battery problems. The data of these four sensors were linearly interpolated between neighbouring sensors. Between 2300 and 3800 m three single point current meters were attached in the cable, every 750 m. The current meters sampled once every 900 s, the ADCP once every 1800 s and the T-sensors once per s.

Typical horizontal current velocities were 0.1 m s$^{-1}$ and tidally dominated. Despite the long mooring line, pressure and tilt sensor information showed generally <1.5° tilt angle and top-buoy excursions across <1.2 m in vertical direction and <100 m in horizontal directions due to current drag. These values were maximally doubled when a Meddy passed above, here between days 350 and 450 (Fig. 2).

The NIOZ3 T-sensors have a precision of <0.0005°C, a response time of $\tau \approx 0.25$ s and a noise level of <0.0001 °C (van Haren et al., 2009). Their clocks were synchronized to within <0.02 s via induction every 4 h.

During the deployment and recovery cruises several shipborne SeaBird 911 CTD-profiles were obtained in the Canary Basin. The CTD-data were used to estimate a temperature-density relationship for quantification of turbulence parameter estimates from the moored T-sensor data. In hindsight, the strong variability with time of water mass properties and the occurrence of intrusions in the area were long considered an obstruction for turbulence parameter estimation from these data.

Attached to the CTD-frame were an upward- and a downward looking 300 kHz ADCP, together forming the 'lowered' LADCP for relative current profiles focusing on vertical current shear and turbulence estimates (e.g., Thomson et al., 1989; Firing and Gordon, 1990). The LADCP data were processed using the inversion method with bottom tracking (Visbeck, 2002; Thurnherr, 2010). The algorithms are available from GEOMAR (Kiel, Germany). Using the shear-scaling of Gregg (1989), turbulence parameters were estimated over 100 m vertical bins.

## 2.1. Estimating turbulence from moored T-sensor data

The T-sensor data were transferred to Conservative Temperature 'Θ' values (IOC, SCOR, IAPSO, 2010) and drift-corrected against a smooth statically stable local mean profile before they were used as a proxy for potential density anomaly $\sigma_1$ referenced to a level of 1000 dBar following

$$\delta\sigma_1 = B\delta\Theta, \tag{1}$$

where B denotes the local slope of the temperature-density relationship. This linear temperature-density relationship was the mean for 300 m of CTD-data around the depth of moored T-sensors, see Section 2.2 for values.

Turbulence parameter estimates of dissipation rate $\varepsilon$ and vertical diffusivity $K_z$ were obtained using the moored T-sensor data by calculating overturning displacement scales after reordering every 1 s the 132.5 m long potential density (re. Θ) profile $\sigma_1(z)$, which may contain inversions, into a stable monotonic profile $\sigma_1(z_s)$ without inversions (Thorpe, 1977). After comparing raw and reordered profiles, displacements

$$d = \min(|z-z_s|)\cdot\text{sgn}(z-z_s), \tag{2}$$

were calculated necessary for generating the stable profile. Then,

$$\varepsilon = 0.64d^2N^3, \tag{3}$$

where N denotes the buoyancy frequency computed from the reordered profile and the constant follows from empirically relating the root-mean-squared (rms) overturning scale with the Ozmidov scale $L_O = 0.8d_{rms}$ (Dillon, 1982), which yields a mean coefficient value from many realizations. Using $K_z = \Gamma\varepsilon N^{-2}$ and a mean mixing efficiency coefficient of $\Gamma = 0.2$ for the conversion of kinetic into potential energy for many realizations of ocean observations (Osborn, 1980; Oakey, 1982; Gregg et al., 2018), we find,

$$K_z = 0.128d^2N. \tag{4}$$

In general, the ocean is a bulk Reynolds number Re environment that has values in the range $10^5 < \text{Re} < 10^7$ that are much larger than in common numerical and laboratory modeling.

According to Thorpe (1977), results from (3) and (4) are only useful after averaging over the size of an overturn. In the following, averaging is applied over at least the vertical scales of the largest overturns and over at least the local buoyancy time scales to warrant a concise mixture of convective- and shear-induced turbulence, and to justify the use of the above mean coefficient values. Here, time-averaging is performed over thousands of profiles over periods of five days, which exceeds the local inertial period. Vertical averaging is over the entire T-sensor range to capture all overturn-scales. Due to the small precision of the T-sensors, thresholds (Galbraith and Kelley, 1996) limit mean turbulence parameter values to $<\varepsilon>_{thres} = O(10^{-12})$ m$^2$s$^{-3}$ and to $<K_z>_{thres} = O(10^{-6})$ m$^2$s$^{-1}$ in weakly stratified waters (van Haren et al., 2015). For the averaging, zeros replace values below threshold. It is verified that using the threshold values instead of zeros for replacement yields different mean values by <5%. This is negligible in comparison with the estimated error of about a factor of two in mean turbulence parameter estimates for the present method (van Haren and Gostiaux, 2012).

In the following, averaging over time is denoted by [...] and averaging over depth-range by <...>. Mean eddy diffusivity values are obtained by averaging the flux first. Standard deviations to the means follow from the averaging.

### 2.2. CTD-LADCP observations

During the LOCO-Canary Basin project, the CTD-LADCP profiling was not aimed at finding and measuring Meddy-contributions. Instead, the profiles were supportive of the background characteristics of internal wave propagation. They also served as a calibration-support for the moored instrumentation. As a consequence, the shipborne profiling provided limited general information on the water property characteristics of the Canary Basin. Examples from around 31±2°N, 23°W during mooring-deployment and -recovery cruises are given in Fig. 3. In these data Meddy-traces, not a Meddy-core, are visible between 900 and 1300 m, above the T-sensor array. In general, within a Meddy or layer of Mediterranean-outflow water salinity increases with depth (e.g., Hebert et al., 1990; Ruddick et al., 2010) so that the temperature-density anomaly relationship coefficient is relatively high: B$_1$ = -

$0.25\pm0.05$ kg m$^{-3}$°C$^{-1}$ for z $\in$ [-1300, -1000] m in the present data. Below a Meddy both salinity and temperature decrease with depth, so that they partially counteract eachother in density contribution. This results in a relatively low absolute value of the temperature-density relationship: B$_2$ = -0.08$\pm$0.02 kg m$^{-3}$°C$^{-1}$ for z $\in$ [-1600, -1300] m in the present data with substantial variability (Fig. 3e). This variability emphasizes the difficulty in using the present T-sensor data as proxies for density variations and, hence, in quantifying turbulence parameters. While the common error in mean turbulence estimates from moored T-sensor data is to within a factor of 2 (van Haren and Gostiaux, 2012), it is about a factor of 3 to 4 for the present dissipation rates due to the relatively large uncertainty in B. Noting that B $\propto$ N$^2$, ε is much more affected than K$_z$. Here somewhat conservatively B = B$_2$ is used, also for the period when a Meddy and hence its core pass close to the mooring with N being nearly twice the value found away in the surrounding waters.

### 2.3. Shapes of overturns

There are several ways to distinguish genuine turbulence from salinity (over-)compensated intrusions that provide false turbulence estimates. In stratified waters, turbulent overturning cannot last longer than the longest buoyancy period. Intrusions can last longer. Additionally, for better distinguishing genuine turbulent overturns from false ones the overturn-shapes can be investigated in the d-z plane. Van Haren and Gostiaux (2014) provide estimates of the slopes z/d of vertical displacement profiles using models of different idealized overturns. After comparison with multiple CTD-profiles from various ocean areas, these authors found that most genuine overturns do not compare with a purely mechanical solid-body rotation that forms a slope of z/d = ½, but with a half-turn Rankine vortex. The Rankine vortex is a 2D-model of a rotating eddy (vortex) in a viscous fluid. It is popular for modelling atmospheric tornadoes. Its maximum current velocity is at some distance from both the vortex-centre and -edge. When used in the x-z plane, it modifies density in a stable linearly stratified fluid. After integrating the Rankine motion over half a turn, the disturbed density field creates z/d slopes that are characterized by: a) borders strictly along a slope just z/d > 1 and which are formed by the upper and lower parts of the density overturn portion (i.e., the parts generated by the edge of the vortex), and b) a slope ranging between ½ < z/d < 1 for the forced inner vortex part and mainly denoting the mechanical overturn. Thus, the Rankine vortex introduces two slopes, one which more or less aligns with the slope of a solid-body overturn forming the long inner axis of a parallelogram, and the other which aligns with the maximum displacements possible in an overturn and delineating the sloping edges of a parallelogram.

A salinity-compensated intrusion of warm-salty water penetrating colder-fresher water of virtually the same density is modeled by a slightly oblique S-shape in density with diffusion/convection incorporated in the intrusive layer (van Haren and Gostiaux, 2014). The resulting displacement-shape, whether or not deformed by diffusion, heavily favors a slope of z/d >= 1 even for the long inner axis of the parallelogram. It thus distinguishes from the 1D-solid-body and the 2D-Rankine overturns.

## 3. Observations and results

### 3.1. Moored data overview

The passage of the Meddy over the mooring between days 350 and 450 was not only noticed in the mooring motion (Fig. 2a), but also in sub-inertial <0.2 cpd (cycles per day) low-pass filtered current components averaged over the range of T-sensors (Fig. 2b). When the two horizontal current components exceeded the sub-inertial 'background' current velocity variability of |0.02| m s$^{-1}$, a clockwise rotating current was observed starting westward at day 350. This suggested a passage from the northwest for an anti-cyclonic eddy. (As this is ambiguous for a single mooring observation, it can also mean a passage from the southeast for a cyclonic eddy). The Meddy-passage was also observed at 2300 m where speeds were about a third of those in Fig. 2b, but it was not found significantly at 3000 m. The passage was associated with a local doubling of near-inertial horizontal current velocity amplitudes (Fig. 2c) and a doubling of both near-inertial and sub-inertial large-scale shear over the 132.5-m range of T-sensors (Fig. 2d). It is noted that the near-inertial shear has larger values at smaller vertical scales O(10 m).

Around 1450 m, the (lower part of the) Meddy-passage was observed in the T-sensor array as the warmest period in the one year of observations (Fig. 4a). Like the currents, temperature varied during the passage and the actual closest-to-core passage was between days 410 and 450. Before and after this 40-day period, irregular smaller warmer and cooler periods alternated. The apparent high-frequency T-variations were due to propagating internal waves, as will become clear from magnifications below.

Turbulence estimates were made for segments of 5 days of drift-corrected data. Time-depth series of the resulting stratification (N) and turbulence dissipation rate, and time series of the 132.5 m vertically averaged turbulence parameters are given in Fig. 4b,c and Fig. 4d,e respectively. A large discrepancy was seen before day 350 and after day 450 compared to the period in between. Between days 350 and 450, stratification was relatively smooth and more or less gradual (Fig. 4b). During the periods before and after, it was concentrated in two or three layers with near-homogeneous waters in between (Fig. 4b). The transitions between these periods of layering and the period of smooth stratification in between were abrupt, as was also observed in turbulence parameters. Turbulence dissipation rate was apparently high in the near-homogeneous layers, and infrequently occurred in small spots only during the warmest period (Fig. 4c). Vertically averaged values (Fig. 4d) were unrealistically high and exceeded values that have been commonly observed in areas with vigorous internal wave breaking like above large underwater topography where $[<\varepsilon>] = O(10^{-7})$ m$^2$ s$^{-3}$ and $[<K_z>] = 10^{-2}$ m$^2$ s$^{-1}$ (e.g., Polzin et al., 1997; van Haren and Gostiaux, 2012). The present larger values were induced by intrusions as will be demonstrated below. It is noted that the largest unrealistic 'false' turbulence values were observed closest to the transitions. The sharp decrease in values during the warm water passage yielded mean values of roughly $[<\varepsilon>] = O(10^{-10})$ m$^2$ s$^{-3}$ and $[<K_z>] \approx 10^{-5}$ m$^2$ s$^{-1}$. These are considered below as representative of genuine ocean-interior turbulence.

*3.2. Transition from intrusion to non-intrusion*

Fig. 5 shows a four-day, 132.5-m vertical range magnification that demonstrates the transition from an intrusion dominated period with false turbulence estimates to an, interrupted, passage of more smoothly stratified waters with genuine turbulence overturns. The main transition was at day 352.5, preceded by a brief transition around day 351.9, with relatively warmer waters before. Whilst internal waves propagated uninterruptedly through the entire image, isotherms were more irregular in the first half of the image (Fig. 5a). This was when large persistent intrusions were observed, e.g. around z = -1420 and around z = - 1500 m, and the reordered profiles demonstrated (temperature-only inferred!) stratification that varied from high values in thin layers to negligibly small in near-homogeneous layers. While such layering still occasionally occurred in the second half of the displayed period, N-extremes were generally smaller (Fig. 5b). Perhaps the transition was most clearly observed in the estimated turbulence dissipation rate (Fig. 5c). Up to day 352.5, large 20 to 40 m tall apparent overturns caused by intrusions dominated four-day, 132.5-m averages $[<\varepsilon>] \approx O(10^{-6})$ $m^2$ $s^{-3}$, which are unrealistic values for the ocean interior. Obviously, the temperature-density relationship was not well established for this period.

This was confirmed from investigating single profiles of these data for their overturn depth-shape slope z/d. From the first two sets of profiles in Fig. 6 it was established that many cores of zig-zag patterns slope largely along z/d = 1, thus not representing a Rankine vortex. In the third set of profiles however, a tiny non-zero turbulence value around z = -1460 m demonstrated a tight z/d = ½ for the overturn core. While this strictly confirms a solid-body rotation model (van Haren and Gostiaux, 2014), the poor vertical resolution prevents distinction from the Rankine vortex model for this small overturn. It lasted about 1.5 h commensurate the local buoyancy period. It is noted that during these first days after transition still small intrusions occurred, which could be traced via the duration of their apparent overturning lasting longer than the local buoyancy period, besides via their overturn-shape slopes.

The departure of the Meddy from the mooring array was equally abrupt (Fig. 7). A few hours before day 456 intrusions became visible around z = -1400 and z = -1500 m, with a sudden transition to thick and persistent intrusion-like layering from day 456 onward, initially mainly around -1460 m.

### 3.3. Ocean-interior episodic turbulent overturning

Fig. 8 shows a similar magnification as Figs 5,7 without any apparent overturning due to intrusions. The temperature time-depth series (Fig. 8a) were again dominated by internal wave motions throughout, with short-scale ones like around day 422.2 superposed on large 50 m crest-trough tidal variations. These isotherm excursions were at least ten times larger than mooring motion excursions. Time series of the latter are shown here around z = -1410 m for reference. All internal wave variations were quite uniform over the 132.5-m vertical range, but did not necessarily display a sinusoidal shape of a purely linear interfacial motion. Generally, isotherms were smooth and only after detailed inspection irregular shapes (with small overturns) were detectable, e.g. around day 421.1, z = -1460 m. The isotherms were about twice as densely spaced compared with the second half of Fig. 5a. As a result, the average stratification was larger here (Fig. 8b), when we adopt the same temperature-density relationship. While the extremities were less compared with the first half of Fig. 5b, it was observed that layering also occurred of alternating strong and weaker stratification.

With the notion that none of the periods demonstrating non-zero overturns lasted longer than the local buoyancy period, the estimated turbulence dissipation rates were sparsely distributed in time-depth (Fig. 8c). It was a puff here and a puff there, resulting in 4-day, 132.5-m averaged values of $[<\varepsilon>] = 4\pm3\times10^{-10}$ m$^2$ s$^{-3}$ and $[<K_z>] = 8\pm3\times10^{-6}$ m$^2$ s$^{-1}$, for $[<N>]$ = $3.2\pm0.5\times10^{-3}$ s$^{-1}$. The mean dissipation rate and eddy diffusivity were similar to within error bounds as the ones estimated using LADCP/CTD for the same depth range, see Fig. 3d for $<\varepsilon>$. They were also similar to within error bounds to microstructure profiler estimates of ocean-interior turbulence (e.g., Gregg, 1989; Polzin et al., 1997).

*3.4. A spectral view*

The uninterrupted three-week period of most stable and smooth stratification, which was assumed to be the close passage of a Meddy-core above, the 'M-period', was compared spectrally with an arbitrary period of equal length during intrusion dominance, the 'I-period' (Fig. 9). Both vertical currents and temperature showed distinctly different internal wave band features. The 130-m vertical mean kinetic energy and the vertical current difference (~shear) spectra demonstrated a significantly larger near-inertial peak during the M- than during the I-period (Fig. 9a). This was already noted in the band-pass filtered time-series data (Fig. 2c). Also, the near-inertial peaks were at slightly sub-inertial frequencies of ~0.97f during the M-period and at slightly super-inertial frequencies of ~1.03f during the I-period. While the kinetic energy of semidiurnal tides was greater during the I-period, the shear in the tidal motions for both the M- and I-periods was not significantly different than that of the background motions. The latter implies shorter vertical length-scales at near-inertial frequencies than at tidal frequencies.

In T-spectra, an f-peak was absent during the I-period, as is common for ocean-interior observations (Fig. 9b). During the M-period however, a significant f-peak was observed in T-spectra with variance equivalent to that of internal tides. Such a near-inertial peak was not commensurate with linear internal wave theory under the traditional approximation that predicts near-horizontal, non-vertical motions at f when locally generated in an environment with zero relative vorticity (e.g., LeBlond and Mysak, 1978). The two T-spectra further deviated in the weaker tidal harmonic peaks during the I-period, when the kinetic energy (Fig. 9a) showed larger peaks at these frequencies. This confirmed that advection or mooring motion by horizontal currents did not dominate the observed T-signals. Higher harmonic fourth-diurnal motions, a common feature of a rectilinear tidal current displacing a mooring, were rather reflecting nonlinearity of the internal tides as observed in the non-sinusoidal isotherm displacements of Fig. 8a. The slope of the 'background' T-spectra outside tidal and near-inertial peaks was also different between the two periods. Noting that the T-spectra in Fig. 9b were scaled with power law $\sigma^{-5/3}$, the frequency $\sigma$ slope (of -5/3 in a log-log fashion plot) representing the inertial subrange of turbulence (Tennekes and Lumley, 1972), both spectra followed this slope around about 10 cpd. This subrange included the minimum buoyancy frequency at about 7 cpd. The M-period-spectrum more or less continued the -5/3-slope, in particular in the internal wave band portion for about $2.1 < \sigma < 13$ cpd. Between 18 and 30 cpd it also showed the same slope, but after a jump in variance. The small deviation was a slow increase, with a net-slope of about -4/3. In the [18, 30] cpd frequency range the apparent spectral noise was far exceeding the near-random noise of statistical significance. It reflected a mix of coherent (wave) and incoherent (turbulence) motions. The apparent inertial subrange slope within the internal wave band was unexpected, certainly for the M-period when smooth stratification and abundant internal wave propagation existed. During the I-period, the slope of the spectral background outside tidal, inertial harmonic peaks was steadily upward for $1.4 < \sigma < 20$ cpd, which resulted in a net-slope of about -1. This slope was considered to be typical for intermittent ocean-interior internal waves (van Haren and Gostiaux, 2009). During both periods, the slope was significantly different from -2, which was previously considered typical for internal waves (Garrett and Munk, 1972; Pinkel, 1981).

The coherence between vertically separated sensors was significantly high for most internal wave band frequencies, for both periods. Between the periods however, it varied for given separation distances $\Delta z$: During the I-period coherence was lost for $\Delta z > 25$ m, except at semidiurnal frequencies (Fig. 9c). During the M-period (Fig. 9d), the internal wave band coherence remained significant for $\Delta z > 40$ m, especially also at the inertial frequency, and at about 13 cpd where the T-variance dipped before suddenly increasing. A small dip was also observed around 7 cpd $\approx 4f_h$, the minimum buoyancy frequency, in Fig. 9d, where Fig. 9c showed a sub-peak. Here $f_h$ denotes the horizontal component of the Coriolis parameter which becomes dynamically important in weak stratification (LeBlond and Mysak, 1978). Around N, coherence dropped steeply into insignificant values, but the spectra remained much smoother during the M-period than during the I-period. While for $\sigma > N$ the latter was irregular (Fig. 9c), the M-period (Fig. 9d) showed a slight increase at about 100 cpd near the maximum small-scale buoyancy frequency $N_{max}$ found in thin layers <10 m thick.

This $N_{max}$ was about equivalent to the mean 10-m shear estimated from the moored ADCP-data during the Meddy-passage. However, this shear-estimate was dominated by random noise. The band-pass filtered mean near-inertial shear-magnitude $|\mathbf{S}|$ was about $2\times10^{-3}$ $s^{-1}$ with peak-values up to about $4\times10^{-3}$ $s^{-1}$. With sub-inertial shear added, the values were 2.5 and $5\times10^{-3}$ $s^{-1}$ for mean and peak, respectively. These values were similar to the mean value of $N = 3.2\times10^{-3}$ $s^{-1}$ for this period. As a result, the bulk Richardson number $Ri = N^2/|\mathbf{S}|^2$ was about unity on 10-m scales, as far as could be established.

## 4. Discussion and conclusions

The estimated turbulence dissipation rates were comparable within one standard deviation with ocean-interior turbulence estimates from present shipborne LADCP/CTD and previous microstructure profiler estimates of (e.g., Gregg, 1989; Polzin et al., 1997). Considering the observed conditions from the present moored observations of near-inertial shear and inertial subrange dominance in the internal wave band, the estimated turbulent overturning is predominantly driven by episodic internal wave breaking. The observed abrupt drop in coherence for $\sigma > N$ suggests a lack of, or across a narrow frequency range, transition from stratified quasi-2D turbulence to fully 3D isotropic turbulence. For isotropic turbulence coherence is expected to be insignificant. However, such a transition could be present in the internal wave band between $N_{min}$ and $N$, which would partition the coherence spectra between shear-induced stratified turbulence and internal waves. This requires future modelling.

While this turbulence was two-three orders of magnitude smaller than observed over steep, large underwater topography (e.g., Polzin et al., 1997; van Haren and Gostiaux, 2012), episodic ocean-interior internal wave breaking does require some nonlinearity. As observed in the time-depth series, the internal wave shapes all had some level of nonlinearity following interactions, which were assumed to be primarily between near-inertial shear and higher-frequency internal waves up to the buoyancy frequency. While most turbulent overturning occurred at vertical scales of <5 m, it was found in or near the weaker stratified layering due to isothermal straining. These overturning sizes are highly comparable with those observed by Gregg (1980) in the Pacific gyre. That area was stable to double diffusion and $K_z = O(10^{-5})$ m$^2$ s$^{-1}$ was mainly due to near-inertial shear instabilities rather than convection, according to the ocean-interior internal-wave-breaking model of Thorpe (2010).

It remains to be investigated whether the present turbulence estimates are representative for the open-ocean-interior. It is noted that they are consistent with previous microstructure profiler estimates except perhaps for an only twice smaller eddy diffusivity. It also remains to be investigated whether the suggested trapping of near-inertial waves underneath a Meddy dominates shear-induced turbulence. The ocean-interior dominance of inertial shear over tidal shear is relatively common. However, a near-inertial peak in T-variance is rarely observed. This peak may be related with the low-frequency vorticity of about -0.05f in the anti-cyclonic Meddy (Pingree and Le Cann, 1993; Bashmachnikov et al., 2009), which is commensurate with the observed peak-shift to sub-inertial frequencies. Such a shift may also be attributed to the weakest stratification $N_{min} = 4f_h$ that provides 0.97f as the lower bound of the inertio-gravity wave band (LeBlond and Mysak, 1978; van Haren and Gostiaux, 2009). Both mechanisms could explain the increase in near-inertial variance and low-frequency peak-shift in shear and kinetic energy. However, a similar peak in T(f)-variance is more difficult to explain. A T(f)-peak requires the ellipses of horizontal near-inertial current velocity paths to deviate from purely circular (in the plane perpendicular to gravity), so that they become elliptic and the vertical current component is non-negligible at f. While this is possible in weakly stratified waters (van Haren and Millot, 2005), it is not known to exist for near-inertial waves trapped in a low-frequency vorticity environment (Kunze, 1985). This requires further theoretical investigation.

The absence of near-inertial T-variance during the I-period either suggests a lack of low-frequency negative (anti-cyclonic) vorticity, or higher minimum buoyancy frequency than during the M-period. While the former may be possible, the latter is difficult to establish from the intrusion-dominated moored observations. While small-scale temperature-variance is high in intrusions, turbulent kinetic energy is found to be low (Oakey, 1988). The present CTD-observations do not show particularly enhanced small-scale variance in density profiles.

The present moored high-resolution T-observations gave some insight in the occurrence of ocean-interior internal wave breaking under predominantly near-inertial shear. While the 2.5-m vertical separation distance did not allow a detailed study of the overturning, the limited shapes and spectral information suggested shear-induced turbulence to dominate over convection. With the 10-m, three-week mean near-inertial shear and associated stratification, the gradient Richardson number of unity suggested a marginally stable ocean, like other stratified environments, e.g. in shelf seas. Future improvements on moored observations seem to be difficult, as sensor separation resolving the smaller Ozmidov-scales needs to be <0.1 m, which constitutes a very difficult constraint on mooring motions of kilometers-long lines.

**Acknowledgments**

I thank the captain and crew of the R/V Pelagia and MTM-technicians for the overboard operations. M. Laan and L. Gostiaux are thanked for their ever-lasting temperature sensor efforts. The NIOZ T-sensors and the large investment program 'LOCO' were financed in part by the Netherlands Organization for the Advancement of Science, N.W.O.

**Table 1**. Non-exhaustive overview of typical mean turbulence values from various areas in the stratified ocean. Shipborne microstructure profiler is indicated by 'μ-prof', CTD/LADCP by 'ladcp'. Moored high-resolution temperature sensor array is indicated by 'moorT'. MAR denotes the Mid-Atlantic Ridge, Smnt a seamount and mab = meters above the bottom. [1]: 14 days averages; [2]: one-hour average.

| Means | ocean-area | $<\varepsilon>$ $(m^2\ s^{-3})$ | $<K_z>$ $(m^2\ s^{-1})$ | Reference |
|---|---|---|---|---|
| μ-prof | thermocline | $10^{-10}$ | $10^{-5}$ | Gregg (1980), Gregg (1989) |
| μ-prof | ocean-interior | $10^{-10}$-$10^{-9}$ | $1$-$3\times10^{-5}$ | Gregg (1989), Polzin et al. (1997) |
| μ-prof | MAR<500 mab | $10^{-7}$ | $10^{-3}$-$10^{-2}$ | Gregg (1989), Polzin et al. (1997) |
| ladcp | Smnt<200mab | $0.5$-$1\times10^{-7}$ | $10^{-3}$ | van Haren and Gostiaux (2012) |
| moorT | Smnt<50mab[1] | $1.5\times10^{-7}$ | $3\times10^{-3}$ | van Haren and Gostiaux (2012) |
| moorT | Bore[2] | $10^{-6}$-$10^{-5}$ | $10^{-2}$ | van Haren and Gostiaux (2012) |
| μ-prof | upper ocean | $8\times10^{-9}$-$5\times10^{-7}$ | -- | Oakey (1982) |

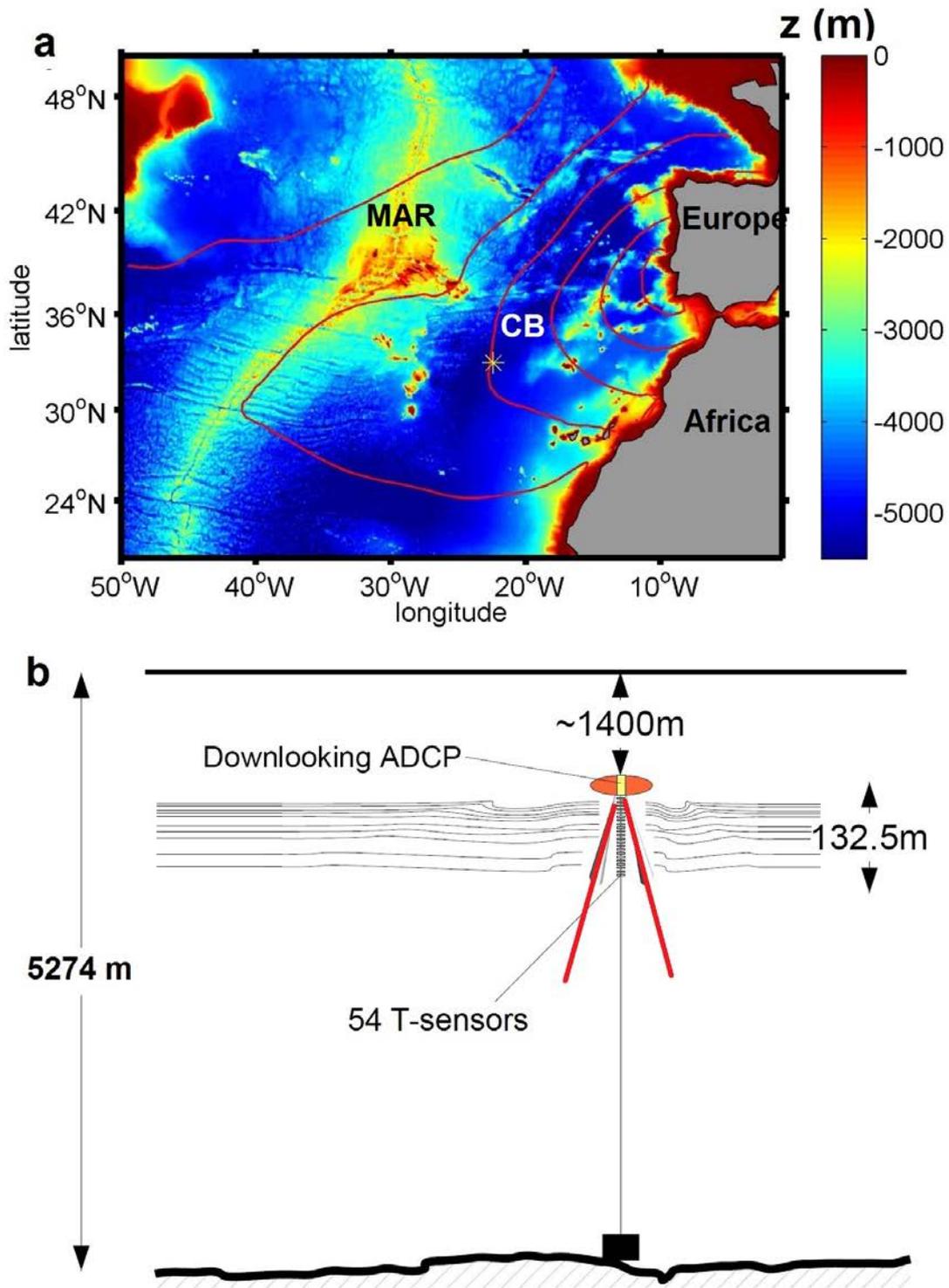

**Figure 1**. Mooring. (a) Site (asterisk) in the Canary Basin 'CB' of the NE-Atlantic, with annual mean salinity contours at 1400 m from World Ocean Atlas 2005 (Antonov et al., 2006). MAR denotes Mid-Atlantic Ridge. Individual Meddies have approximately the horizontal size of the asterisk. (b) Configuration.

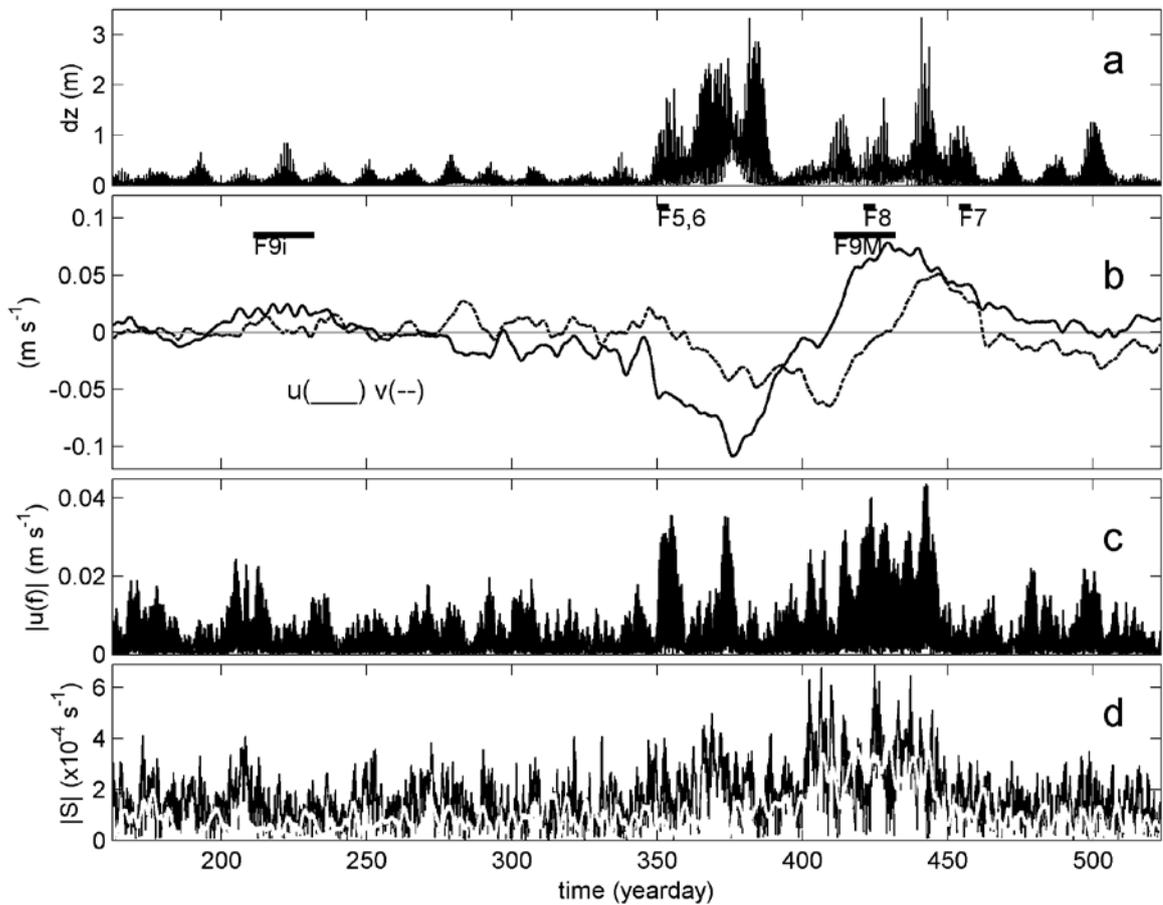

**Figure 2**. Current velocities and mooring motion, for the first year of data. (a) Vertical top-buoy displacement due to current drag as estimated from tilt and pressure information. (b) Horizontal current components averaged over the range between 1400 and 1530 m and low-pass filtered using a phase-preserving double-elliptic filter with ripple suppression and cut-off at 0.2 cpd. Solid graph is u- (positive Eastward), dashed graph v- (positive Northward) component. The periods of Figs 5-6, 7, 8 and 9 are indicated with 'F' and number. (c) Magnitude of band-pass filtered near-inertial u-component. (d) Shear-magnitude over the 132 m range of T-sensors from filtered ADCP-data. A near-inertial band-pass filter yields the solid graph and a 0.2 cpd cut-off low-pass filter the white graph.

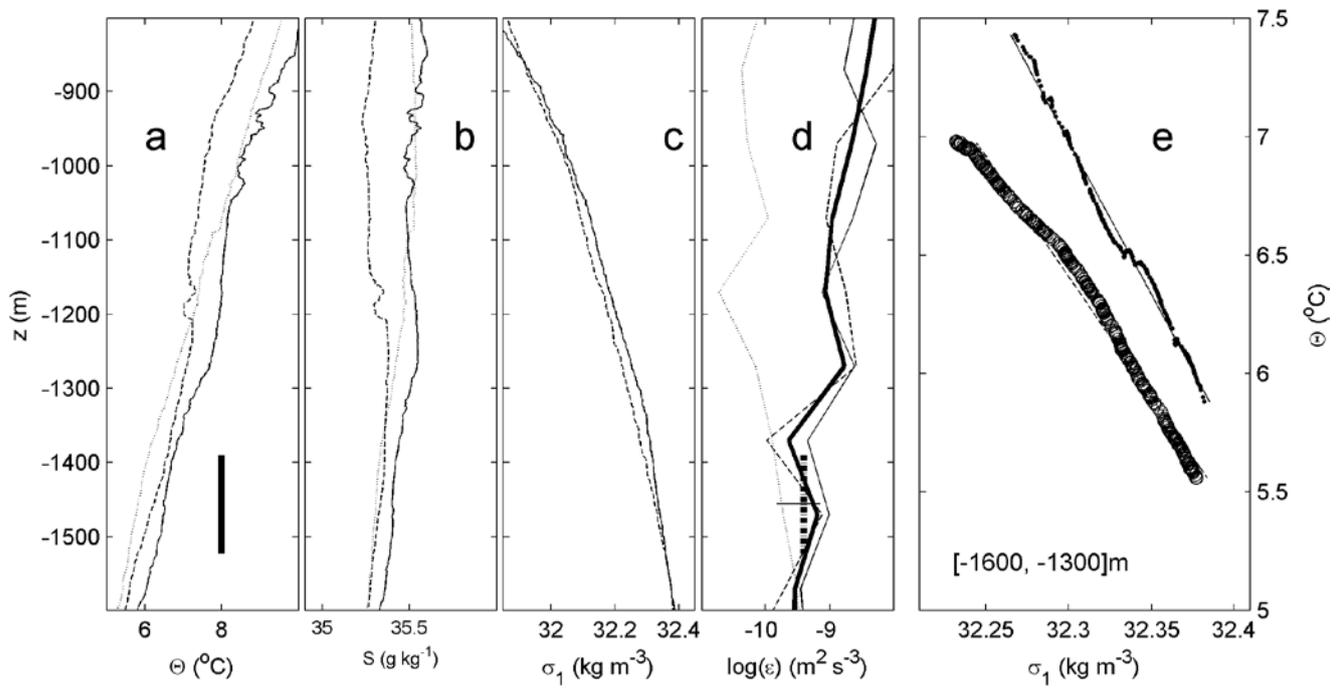

**Figure 3**. Excerpt from shipborne CTD- and LADCP-profiles in the Canary Basin during deployment cruise in 2006 (solid graphs) and recovery cruise in 2007 (dashed graphs). Additionally, in some panels a profile from 2004-data is shown (dotted). (a) Conservative Temperature. The vertical bar indicates the depth range of moored T-sensors. (b) Absolute Salinity. The x-axis range is similar to the one of a. in terms of density contribution. (c) Density anomaly referenced to 1000 dbar. (d) Logarithm of dissipation rate from 100 m vertical bin-averaged LADCP profiles. The mean of these profiles is the thick-solid graph. The dashed vertical bar indicates the time-depth mean (±1 sd, standard deviation) of mooring data in Fig. 8c. (e) Relationship between the data in a. and c. for the indicated depth range.

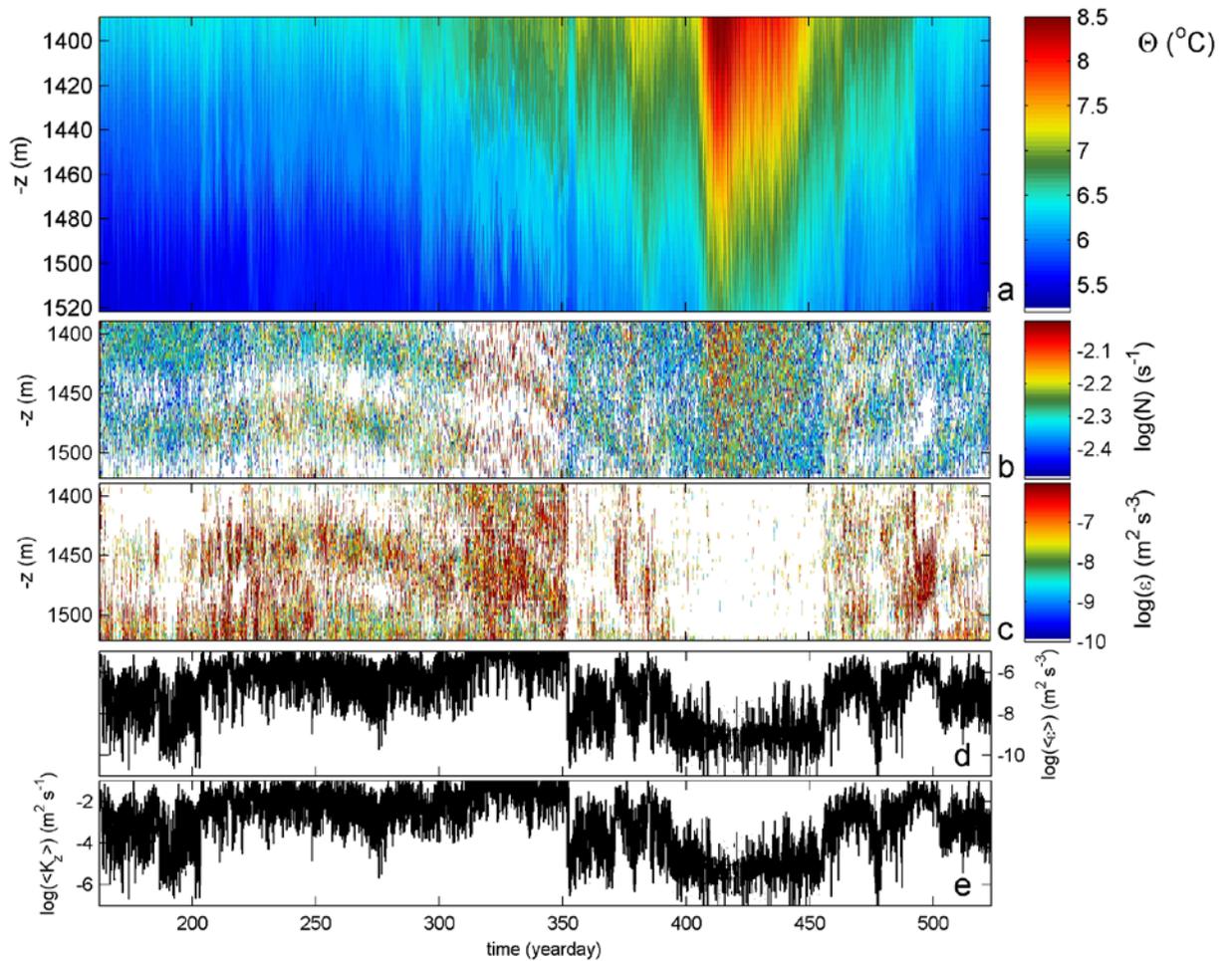

**Figure 4**. Same one year period as in Fig. 2, here providing an overview of T-sensor data. The passage of a Meddy above is noticeable in the discontinuities around days 352.5 and 456. (a) Conservative (~potential) Temperature, with data from four sensors interpolated. (b) Logarithm of buoyancy frequency from the reordered version of a. (c) Logarithm of dissipation rate estimates from data in a., b. (see text). (d) Vertically averaged data of c. (e) Logarithm of vertically averaged eddy diffusivity.

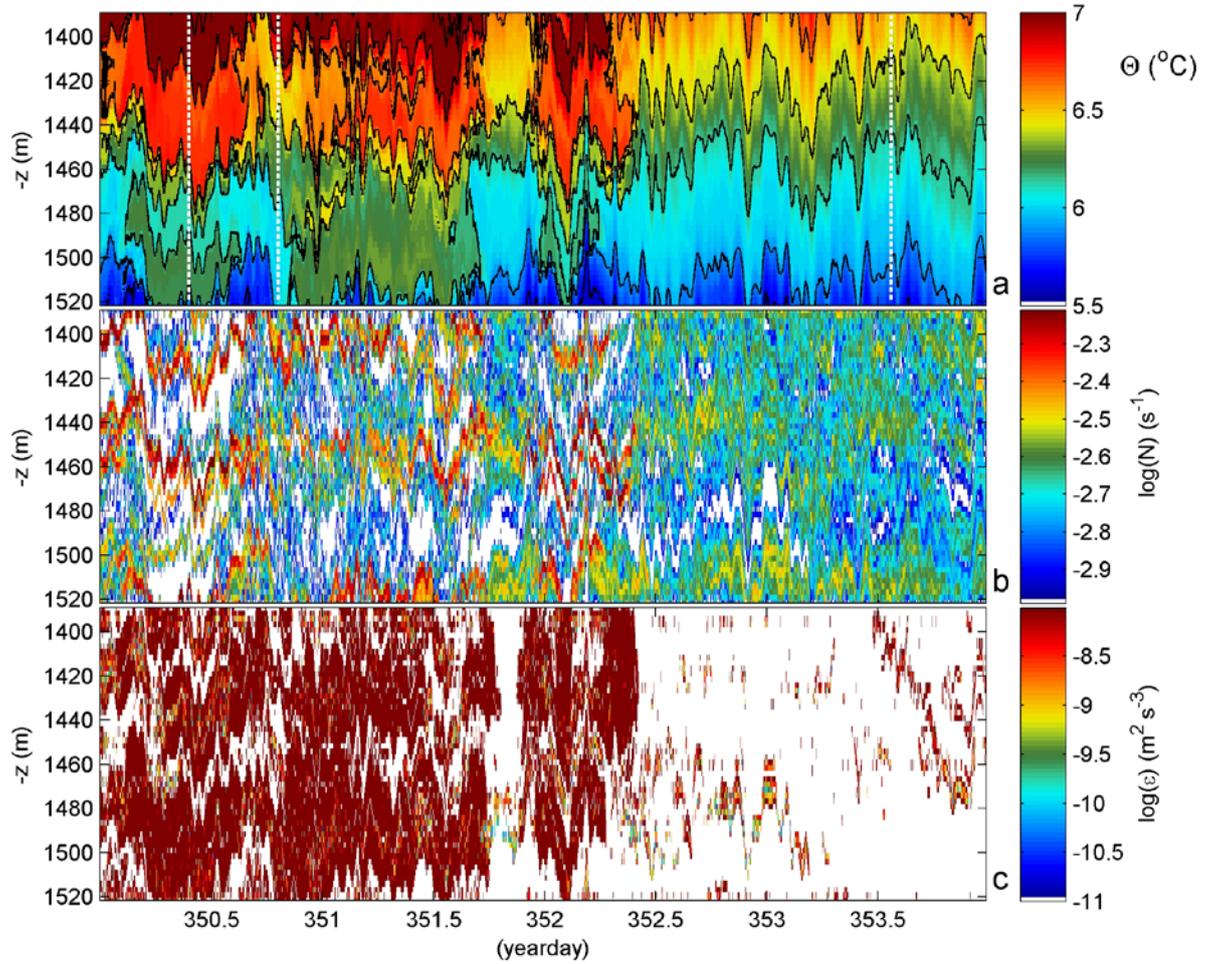

**Figure 5**. Four day time-depth series of T-sensor data around the transition between intrusion-dominated step-like stratification up to day 352.5 and more gradual stratification after day 352.5 assumed to be related with the passage of a Meddy. (a) Conservative Temperature with black contours drawn at 0.25°C increments and the vertical white-dashed lines indicating the times of profiles of Fig. 6. (b) As in Fig. 4b. (c) As in Fig. 4c.

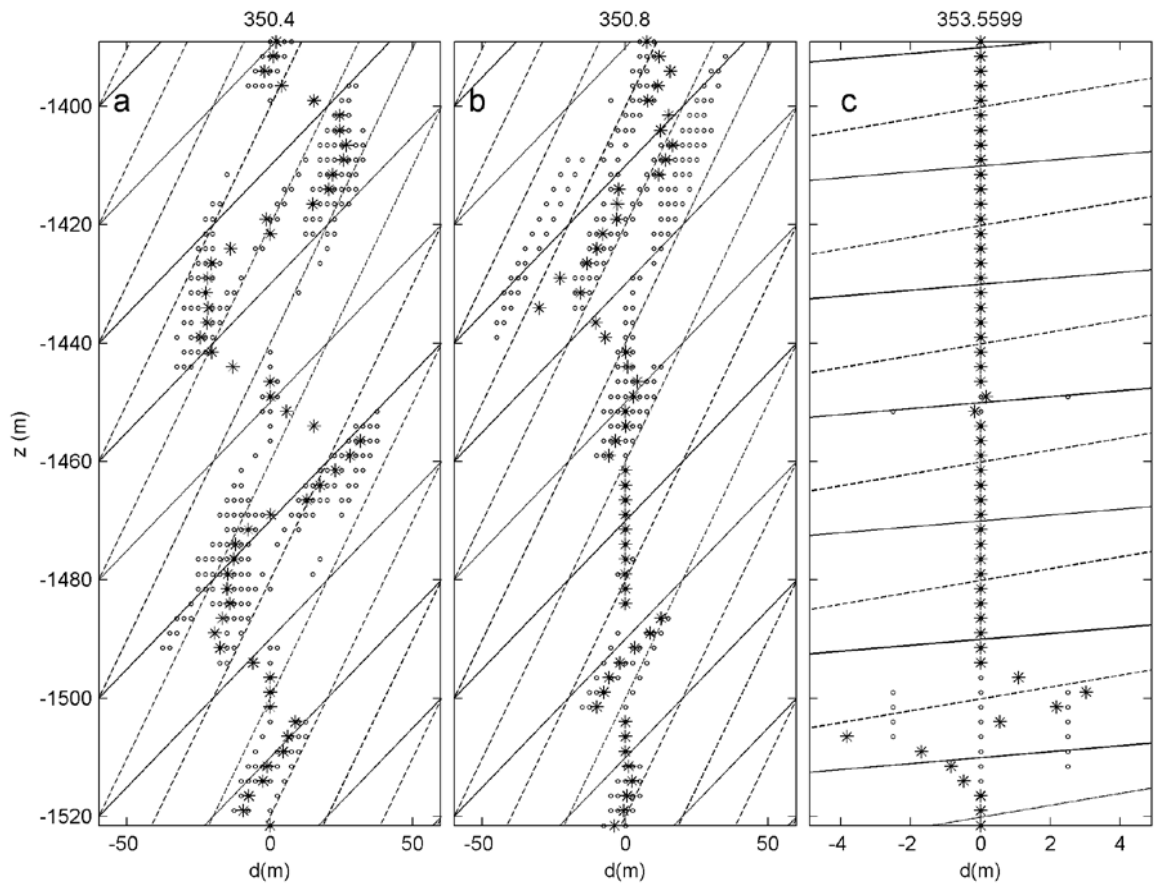

**Figure 6**. Three examples of ten consecutive profiles of displacements d (o) and their mean values at given depths (*) for times indicated above and along the vertical white-dashed lines in Fig. 5a. The grid consists of two slopes in the d-z plane: $z/d = \frac{1}{2}$ (solid lines) and $z/d = 1$ (dashed lines). For the method see Section 2.3 and further in van Haren and Gostiaux (2014).

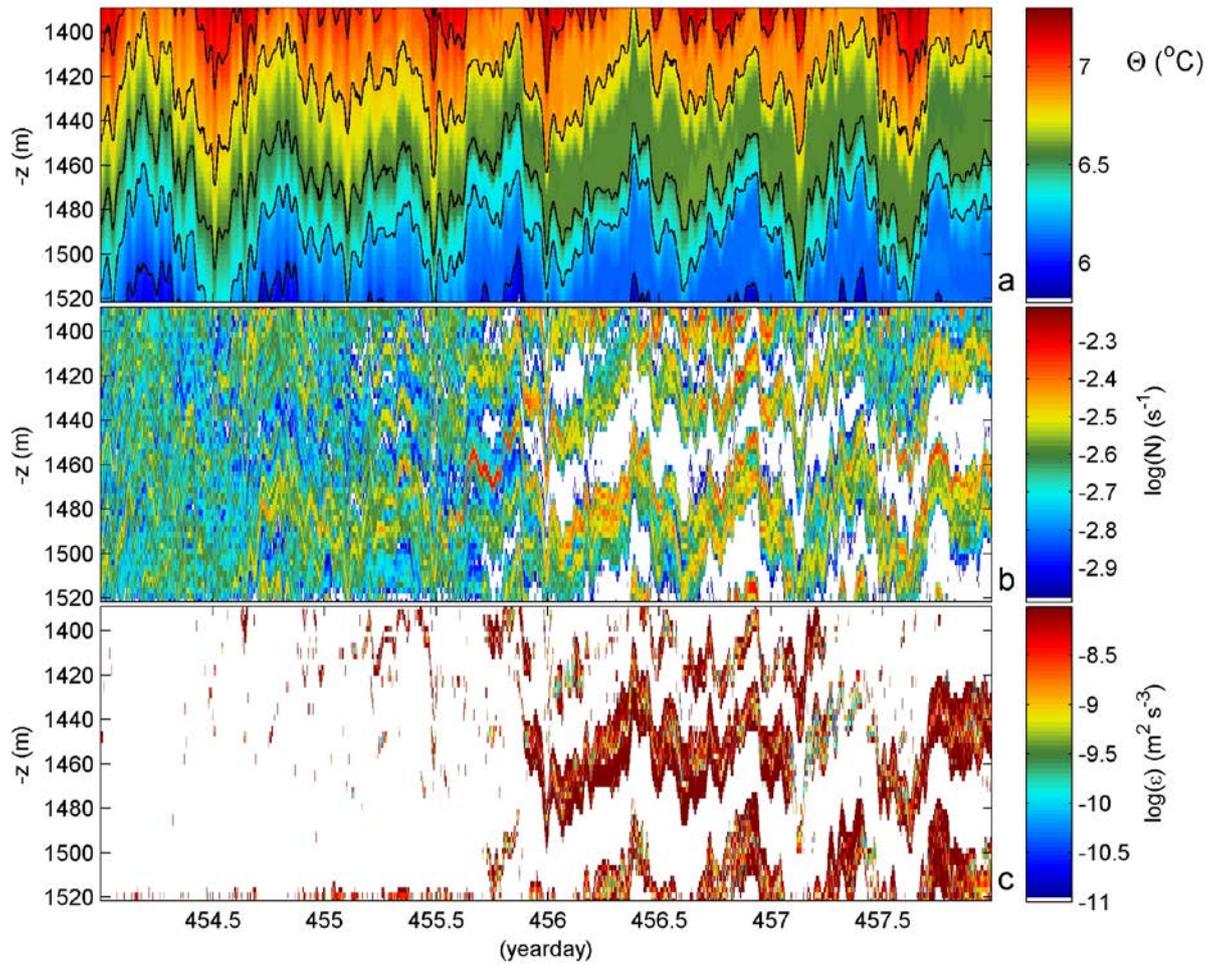

**Figure 7**. Same as Fig. 5, but around day 456 of the Meddy departing. Note the different temperature range in a.

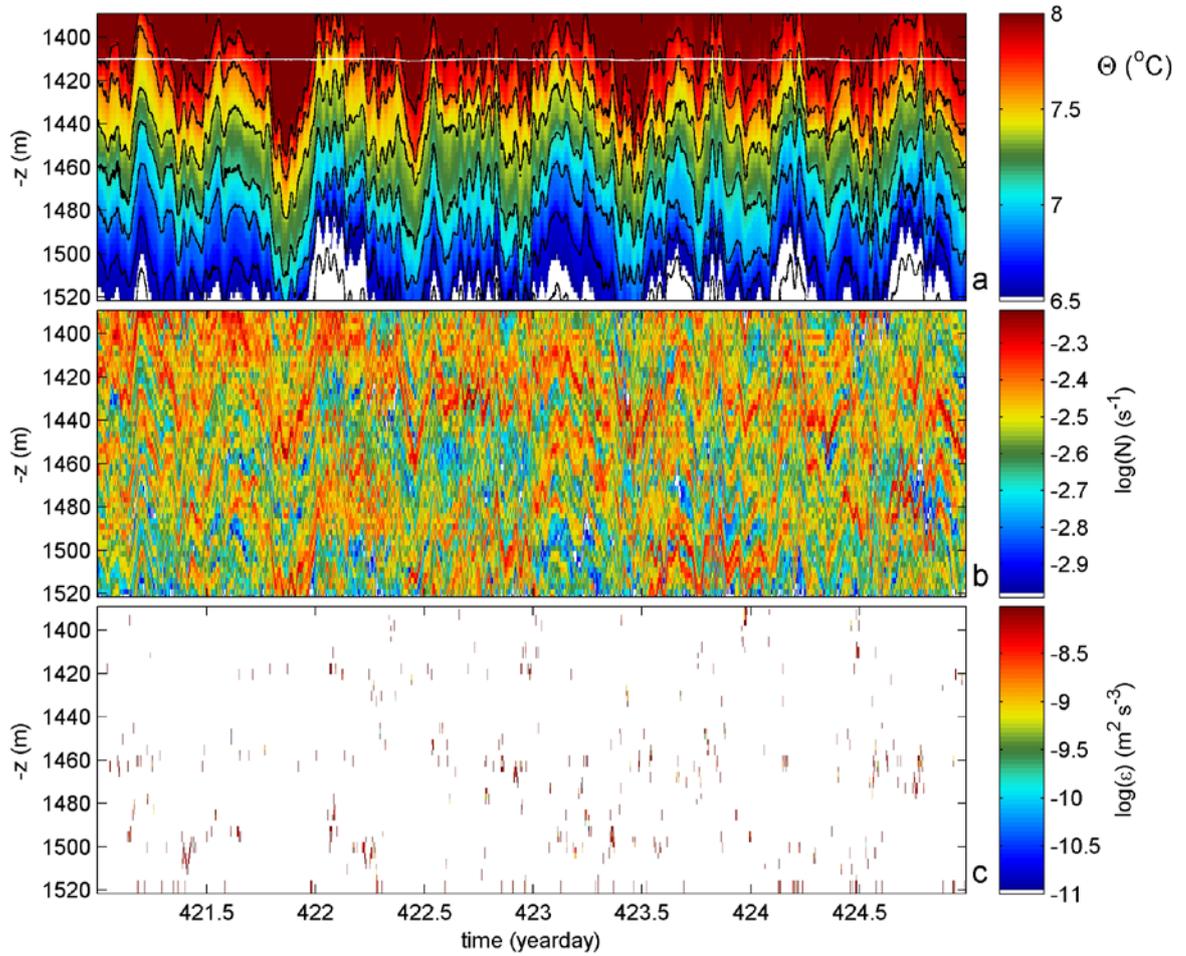

**Figure 8**. Same as Fig. 5, but during the passage of a Meddy. Note the different temperature range in a. The thin white line around -1410 m represents the time-series of the top-buoy vertical displacements of Fig. 2a.

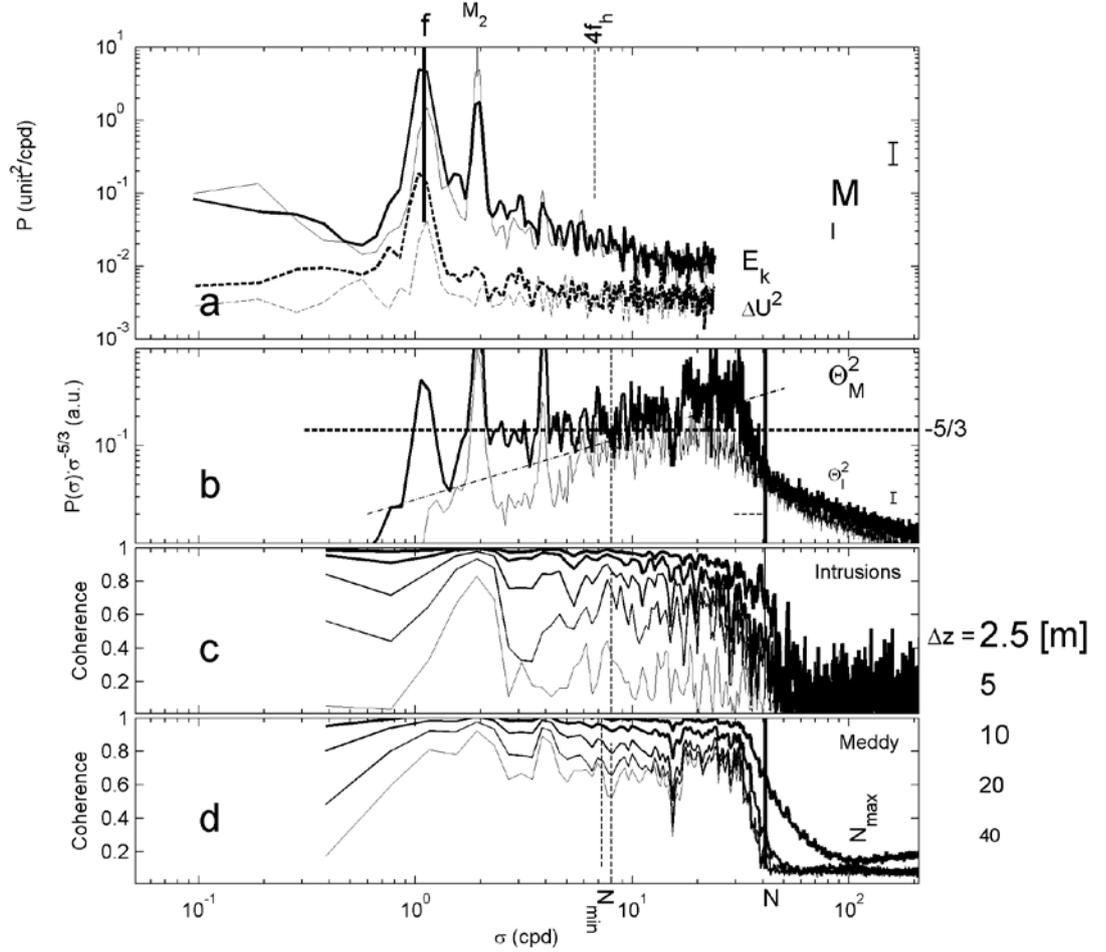

**Figure 9**. Spectral overview for the periods between days 211 and 232, representing a period dominated by intrusions 'I' (thin graphs in a., b.), and between days 411 and 432, representing a Meddy 'M' passage above (bold graphs in a., b.). (a) Moderately smoothed (30 degrees of freedom, dof) spectra of kinetic energy from ADCP averaged over vertical bins 1-14 (solid), which corresponds with the range of T-sensors, and current velocity difference (between the depths of lower and upper T-sensor; dashed). The local inertial frequency f is indicated together with the semidiurnal tidal frequency $M_2$ and the local minimum buoyancy frequency $N_{min} \approx 4f_h$, $f_h$ denoting the horizontal Coriolis parameter. (b) Strongly smoothed ($\sim$100 dof) spectra of 20-s sub-sampled T-data representing M- and I-periods. Spectra are scaled with the inertial subrange slope of $\sigma^{-5/3}$, which is a flat horizontal line as in the dashes for reference. The dash-dot line slopes as +2/3 in this frame, representing a net spectral decay sloping like $\sigma^{-1}$. The vertical dashed line indicates $N_{min}$ and the solid vertical line the mean buoyancy frequency N during M-period, which may have a frequency extent as indicated by the thin horizontal dashed line during the I-period. (c) Coherence spectra for the I-period for all T-sensor-pairs at the separation distances indicated to the right (larger symbol = thicker line). (d) As c., but for the M-period.